\documentstyle[12pt]{article}

\newcommand{\D}{{\rm d}}
\newcommand{\r}{{\bf r}}
\newcommand{\s}{{\bf s}}

\begin{document}


\topmargin 0pt
\renewcommand{\thefootnote}{\fnsymbol{footnote}}
\newpage
\setcounter{page}{0}

\begin{titlepage}

\begin{flushright}
November 1994\\
cond-mat/9411126
\end{flushright}

\vspace*{1.0cm}

\begin{center}
{\large \bf FERMIONS IN A RANDOM MEDIUM} \\

\vspace{1.5cm}
\setcounter{footnote}{0}
Harald Kinzelbach$^{(1)}$\footnote{
Electronic mail: kibach@iff078.iff.kfa-juelich.de}
\setcounter{footnote}{6}
and Michael L\"assig$^{(2,1)}$\footnote{
Electronic mail: lassig@iff011.dnet.kfa-juelich.de }  \\

\vspace{1.0cm}
${}^{(1)}${\sl Institut f\"ur Festk\"orperforschung, Forschungszentrum,
               52425 J\"ulich, Germany}   \\
${}^{(2)}${\sl Max-Planck-Institut f\"ur Kolloid- und Grenzfl\"achenforschung,
               14513 Teltow, Germany}

\end{center}
\vspace{1.5cm}
\setcounter{footnote}{0}

\begin{abstract}

We study the continuum field theory for an ensemble of directed polymers  $\r_i
(t)$ in $1+d'$  dimensions that live in a medium with quenched point disorder
and  interact via short-ranged pair forces $  g \Psi (\r_i - \r_j) $. In the
strong-disorder (or low-temperature) regime, such forces are found to be
relevant in any dimension~$d'$ below the upper critical dimension for a
single line.
Attractive forces generate a bound state with localization length
$\xi_\perp \sim |g|^{- \nu_\perp}$; repulsive forces lead to mutual avoidance
with a pair distribution function
${\cal P}(\r_i - \r_j) \sim |\r_i - \r_j|^\theta$ reminiscent of
interacting fermions. In the experimentally important dimension $d' = 2$, we
obtain  $\nu_\perp \approx 0.8$ and $\theta \approx 2.4$.

\end{abstract}

\end {titlepage}


\renewcommand{\thefootnote}{\arabic{footnote}}

Flux lines in dirty type-II superconductors  \cite{fluxlines} are a well-known
example of  low-dimensional manifolds embedded in media with quenched disorder
(reviewed e.g. in \cite{reviews}). Such systems are of interest
also because of their links to more complicated random systems such as spin
glasses \cite{relation.spinglass}, to surface growth
\cite{KPZ,KrugSpohn.review}, and to randomly driven hydrodynamics
\cite{noisy.Burgers}. A well-studied case is the low-density  limit of a {\em
single} line in a medium with quenched point impurities. The statistical
properties of such a line differ from those of a free, thermally fluctuating
line. Moreover, the disorder modifies the interactions of the line with other
objects. For example, the effect of a rigid line defect (``columnar'' defect)
parallel to the preferred axis of the lines turns out to be {\em weaker} than
in a pure system. A weakly attractive defect localizes the line only up to the
borderline dimension $d'_\star = 1$; in higher dimensions, the transition to a
localized state takes place at finite coupling strength
\cite{columnar.defect,DEPINNING}. In a pure system, the borderline dimension is
$d'_\star = 2$ \cite{Lipowsky.fluctuations}.

Mutual interactions between several identical lines in a disordered medium are
the subject of this letter. We study their effects by the methods of
continuum field theory. Related aspects of this system have been treated
by Bethe ansatz methods in $d' = 1$~\cite{KardarNelson.CIT},
on a lattice~\cite{Mezard,Tang}, and in a Wilson renormalization
group~\cite{NattermannAl,Mukherji}, see the discussion below.

In a pure system, weakly attractive short-ranged forces generate a bound state
up to the same dimension $d'_\star = 2$ as do columnar defects.  Unlike in the
case of columnar defects, however, the effect of pair interactions between
identical lines is {\em enhanced} by the disorder, which (at sufficiently low
temperature) leads to a bound state in any dimension where the low-temperature
behavior of a single line is governed by non-thermal scaling exponents. Of
equal importance are repulsive forces; for example, the magnetic interaction
between flux lines can be treated as a contact interaction at sufficiently low
densities \cite{fluxlines,NattermannAl}. In a pure system, such forces are
important only in $d' = 1$, where they act as an effective constraint on the
fluctuations that is equivalent to the Pauli principle: the lines behave like
the worldlines of {\em free fermions} characterized by a pair distribution
function ${\cal P}(\r_i - \r_j) \sim |\r_i - \r_j|^\theta$  with $\theta = 2$
\cite{fermions1,fermions2}. In higher  dimensions, there is no long-ranged
effect on the pair-distribution function. In the presence of disorder, on the
other hand, we find that ${\cal P}(\r_i - \r_j)$ obeys a power law also for $d'
> 1$,  with a new exponent $\theta$ that depends on the dimension. This
behavior is reminiscent of {\em interacting fermions}, which is why we call
these mutually avoiding lines fermions in a random medium.

This  strong effect of repulsive interactions can be understood qualitatively
as resulting from the energetic competition with an effective line-line
attraction due to the impurities. In almost all realizations of the disorder,
there is a unique ground state, i.e. a path of minimal energy
\cite{comment.minpath}. At low temperatures and without direct forces, two
lines will share this path with finite probability even in a system of infinite
size \cite{Mezard,Parisi,HwaFisher.paths}. Repulsive interactions, however,
force one of the lines into a (distant) excited state.

Our calculations in the sequel will be restricted to the case of two
interacting lines in a random medium; the generalization to an arbitrary
number of lines is straightforward.
On a mesoscopic scale, the system is described by an effective continuum
Hamiltonian
\begin{equation}
{\cal H}  =  \sum^2_{i=1} \; \int { {\rm d} t \,
\left\{  \,   \frac{1}{2}   \left(  \frac{ {\rm d} \r_i }{{\rm d} t} \right)^2
- \eta (\r_i,t)  \right \} }  \;
+  \; g \, \int {  {\rm d} t \; \Psi (t)   } \; .
\label{H}
\end{equation}
Here $\r_1(t)$ and $\r_2(t)$  denote the displacement vectors of the two
lines (also called  {\em directed polymers}) in $d'$ transversal
dimensions, as a function of the longitudinal ``timelike'' coordinate $t$.
Both lines are subject to the same random potential $\eta (\r,t)$
which models the quenched point defects.
It is assumed to be Gauss-distributed with
$ \overline{\eta  (\r,t) } = 0 $ and
$ \overline{ \eta (\r,t) \eta (\r',t') } =
2 \sigma^2 \, \delta^{d'} (\r - \r')  \delta (t - t')$.
Averages over disorder are denoted by an overbar, thermal averages
by brackets $ \langle \dots \rangle $. The direct interaction between the
lines, which is assumed to be short-ranged, is described by the continuum field
$\Psi (t) = \delta^{d'} \left(\r_1(t) -  \r_2(t)  \right)$.

For $g = 0$, the two lines feel only the random potential, and the free energy
of the system is just twice the free energy of a single line. The continuum
field theory for this system serves as the starting point for a systematic
renormalized perturbation theory in the line-line coupling $g$. The analysis is
complicated by the fact that even the ``free'' theory ($g = 0$) has
non-Gaussian multipoint correlations due to the quenched averaging.
Nevertheless, we are able to obtain solely in terms of single-line properties
the scaling dimension and the form of the
operator product expansion of the pair interaction field $\Psi (t)$,
see (\ref{x}) and (\ref{OPE}) below.
These determine the  renormalization group  equations for the
interaction strength to leading order, and hence the phase diagram of the
system \cite{thermalRG}.

The large-scale behavior of a single line is generated by the sample-to-sample
fluctuations of the ground state paths and defines two important exponents. The
{\em roughness exponent} $\zeta$ characterizes the mean transversal excursions
of the line, given e.g. by the two-point function $\overline{ \left\langle (
\r(t) - \r(t') )^2  \right\rangle } \sim \vert t - t'  \vert^{2 \zeta} $.  The
exponent $-\omega$ is the anomalous dimension of the {\em disorder-averaged
free  energy}   $\overline F = \beta^{-1} \overline{ \log {\rm Tr} \exp (-\beta
{\cal H}) } $, whose universal part has the scaling form  $\overline F(T,R)
\sim T^\omega {\cal F} (R / T^\zeta)$ in a finite system of  transversal size
$R$ and longitudinal size $T$ \cite{HuseHenley.roughening}. The two exponents
are related by a ``tilt'' symmetry of the system, $\omega = 2 \zeta -1$ (see
e.g. \cite{FisherHuse.paths,KrugSpohn.review}).  In low dimensions, $\zeta$ is
always larger than in the case of thermal fluctuations, namely $\zeta = 2/3$
for $d' =1$ and $\zeta  \approx 0.62$ for $d' =2$. For $d' > 2$, a phase
transition appears at a finite temperature; in the high-temperature phase, the
system is asymptotically thermal, i.e. $\zeta = 1/2$ and $\omega = 0$ (see e.g.
\cite{KrugSpohn.review} and references therein). Whether a finite {\em upper
critical dimension} $d'_>$ exists such that for $d' \geq d'_>$ the thermal
exponents govern also the low-temperature phase is controversial; some workers
believe $d'_> \approx 4$  \cite{MooreAl.modecoupling}. In the continuum theory
(\ref{H}) with $g = 0$, the large-scale regime is reached in a  crossover from
the Gaussian theory with characteristic longitudinal length $\tilde
\xi_{\scriptscriptstyle \|} = \beta (\sigma^2 \beta^3)^{-2/(2-d')}$. We have
discussed the renormalized continuum field theory for this regime in  ref.
\cite{DEPINNING}. Its construction  involves a reparametrization of the
transversal displacement and of the free energy,  $\r \to \beta^{1/2} \tilde
\xi_{\scriptscriptstyle  \|}^{\omega/2} \r$ and $\overline F \to \beta \tilde
\xi_{\scriptscriptstyle
\|}^\omega \overline F$, such that the renormalized variables remain finite in
the limit $\tilde \xi_{\scriptscriptstyle  \|} \to 0$ (i.e. $\beta^{-1} \to 0$
or $\sigma^2
\to \infty$).  The additional reparametrization
$\Psi \to \beta^{- d'/2} \tilde \xi_{\scriptscriptstyle  \|}^{ d'/2} \Psi$
and  $g \to \beta^{1+d'/2} \tilde \xi_{\scriptscriptstyle  \|}^{\omega - d'/2}
g$
keeps all correlation functions of the pair field free of singularities in
this limit, as we will see explicitly below. This is accompanied by a
temperature reparametrization; the renormalized temperature $ \beta^{-1} \sim
\tilde \xi_{\scriptscriptstyle  \|}^\omega $ is an irrelevant scaling variable
of
dimension~$- \omega $.

With this renormalized theory for $g = 0$ at hand, the effect of pair
interactions may now be expanded in powers of the coupling constant $g$.
For example, the perturbation series for the free energy density
$ \overline f \equiv \lim_{T \to \infty} \overline F / T$ in a system of
transversal size $R \equiv L^\zeta$,
\begin{equation}
\overline{f (g, L)} - \overline{f(0, L) } =
- \beta^{-1}   \sum_{N = 1}^\infty
\frac{ (-\beta g)^N }{N!}
\int {\rm d} t_2 \dots {\rm d} t_N \;
\overline{ \left\langle  \Psi (0)  \Psi (t_2) \dots
\Psi (t_N)
\right\rangle^c } \;,
\label{expansion}
\end{equation}
involves integrals over connected correlation functions of the pair field
$\Psi$ evaluated at $g = 0$. We are hence led to study these correlation
functions. While
they are finite by construction, the ``time'' integrations generate new
singularities that have to be handled by an additional renormalization. (The
longitudinal scale $L$ plays the r\^ole of an infrared cutoff and will be used
to parametrize the renormalization group flow.)

The one-point function $\overline{ \left\langle \Psi (t) \right\rangle }$
gives the probability density that the two lines intersect  at time $t$,
averaged over thermal and disorder fluctuations. We will show below that
for $R^{1 / \zeta} \gg \tilde \xi_{\scriptscriptstyle  \|}$,
\begin{equation}
\overline{ \langle \Psi (t) \rangle } \, = b \; R^0    \;,
\label{x}
\end{equation}
where $b$ is a constant independent of $\tilde \xi{\scriptscriptstyle \|}$.
Hence $\Psi (t)$ can be regarded as a scaling field of dimension zero.
Eq.~(\ref{x}) expresses the fact that for $g = 0$, the two lines follow
the same path (of minimal energy) with finite probability even in the
thermodynamic limit $R \to \infty$.

The higher connected correlation functions of the local pair field $\Psi$ can
be
shown to obey the {\em operator algebra}
\begin{equation}
\Psi (t) \Psi (t') = c \beta^{-1} |t - t'|^{-\omega} \, \Psi (t') + \dots
\label{OPE}
\end{equation}
with a  coefficient $c > 0$, which is valid as an asymptotic identity inserted
into any such correlation function  $\overline{ \langle \dots \Psi (t) \Psi
(t') \dots \rangle^c} $ for $ t \to t'$. This type of operator algebra is
familiar from ref.~\cite{DEPINNING} (where the reader is referred to for a more
detailed discussion): the field $\Psi$ couples to itself, but the leading
singularity in (\ref{OPE}) involves a correction-to-scaling exponent related to
the irrelevant coupling constant $\beta^{-1}$.

To establish eq.~(\ref{x}), it is useful to study the pair distribution
function
${\cal P} ({\bf d}) \equiv
  \overline{\langle \delta^{d'} ( \r_1 (t) - \r_2 (t) - {\bf d} ) \rangle }$,
i.e. the probability density that the two lines have the relative displacement
${\bf d}$. We find that for $g = 0$ and in the thermodynamic limit $R \to
\infty$, ${\cal P}$ has a scaling form that depends only on $|{\bf d}|$ and
$\tilde \xi_{\scriptscriptstyle  \|}$ with the asymptotic behavior
\begin{equation}
{\cal P} ({\bf d}) \sim \left\{
\begin{array}{l@{\quad \mbox{for} \quad}l}
\tilde \xi_{\scriptscriptstyle  \|}^\omega \; |{\bf d}|^{- (d' \zeta + \omega )
/ \zeta}
& |{\bf d}| \gg \tilde {\xi_{\scriptscriptstyle \|}}^{\zeta} \;,
\\
\tilde \xi{\scriptscriptstyle \|}^{- d' \zeta} & |{\bf d}| \ll
\tilde {\xi_{\scriptscriptstyle \|}}^{\zeta} \;,
\end{array} \right.
\label{prob}
\end{equation}
see also the extensive discussion in~\cite{HwaFisher.paths}. Recall that
$\tilde \xi_{\scriptscriptstyle  \|}^\omega$ is the renormalized temperature
$\beta^{-1}$.
Hence with finite probability, the two lines share a common ``tube'' of width
$ \xi_{\scriptscriptstyle  \|}^\zeta $ (i.e. $ |{\bf d}| <
\xi_{\scriptscriptstyle  \|}^\zeta $), but they
do make large excursions whose probability decays with a power
of $|{\bf d}|$. To show this, we first rewrite ${\cal P} ({\bf d})$ for $g = 0$
and ${\bf d} \neq 0$ in terms of the properties of a {\em single} line,
\begin{eqnarray}
{\cal P} ({\bf d})
&=& \int {\rm d^{d'}}  {\bf s}  \;\;
\overline{ \langle \delta^{d'} ( \r_1 (t) - {\bf s} ) \;
                   \delta^{d'} ( \r_2 (t) - {\bf s} - {\bf d} )
           \rangle }
\\
& = & \int {\rm d^{d'}}  {\bf s}  \;
\overline{ \langle \Phi (t, {\bf s}) \rangle_1 \;\;
           \langle \Phi (t, {\bf s} + {\bf d})  \rangle_1 }
\\
& = & - \int {\rm d^{d'}}  {\bf s}  \;
\overline{ \langle \Phi (t, {\bf s}) \;
                   \Phi (t, {\bf s} + {\bf d})  \rangle_1^c }  \;.
\label{onepoint1}
\end{eqnarray}
In the single-line system,
$ \Phi (t, {\bf s}) \equiv \delta^{d'} ( \r(t) - {\bf s} ) $, and averages are
marked by the subscript 1. Here we have exploited the fact that without pair
forces, the two lines are independent in any realization of the disorder,
and that the full correlation function
$ \overline{ \langle \Phi (t, {\bf s})
                     \Phi (t, {\bf s} + {\bf d}) \rangle_1 } $
vanishes since a directed line has a single-valued path $r(t)$. Next we use
the operator algebra
$ \Phi (t, {\bf s}) \Phi (t, {\bf s} + {\bf d}) \sim
 - c_1 \beta^{-1} \vert {\bf d} \vert^{-(d' \zeta + \omega) / \zeta}
\Phi (t, {\bf s}) + \dots $ with $c_1 > 0$
(that follows directly from the results of \cite{DEPINNING}
or \cite{HwaFisher.paths}) and the normalization
$ \int \D^{d'} \s \; \overline{ \langle \Phi (t, {\bf s}) \rangle_1 } = 1 $ to
obtain (\ref{prob}) in the regime
$|{\bf d}| \gg \tilde {\xi_{\scriptscriptstyle \|}}^{\zeta}$; the opposite
limit then follows from the normalization
$ \int \D^{d'} {\bf d} \; {\cal P} ({\bf d}) = 1 $
and the relation
$ \int {\rm d^{d'}} {\bf d} \; {\cal P} ({\bf d}) \; {\bf d}^2  =
  2 \beta^{-1} R^{1/\zeta} $,
a consequence of the tilt symmetry \cite{comment.prob}.
For $R \to \infty$, notice that
${\cal P} (\bf d)$ as given by (\ref{prob}) is only normalizable for $\omega >
0$. The expectation value of the pair potential $\Psi$ at $g = 0$ involves an
integral over the potential function times $ {\cal P} ({\bf d}) $. For a
potential of microscopic range $\Delta_0$ which is well-behaved at ${\bf d} =
0$, this
integral is independent of $R$, which is just eq.~(\ref{x}).
For $\Delta_0 \ll \tilde \xi_{\scriptscriptstyle  \|}^\zeta$, the integral
develops a
singularity $ \tilde \xi_{\scriptscriptstyle  \|}^{-d' \zeta}$
that we have absorbed into the
definition of the {\em renormalized} field~$\Psi$.

The form of the operator algebra (\ref{OPE}) could be derived by arguments
very similar to the ones we have used in ref.~\cite{DEPINNING}.
Here we take a different  avenue: we exploit the exact
mapping of the quenched polymer system defined by (\ref{H}) to a nonlinear
stochastic evolution equation, a generalization of the  Kardar-Parisi-Zhang
equation~\cite{KPZ}. If $Z(\r_1, \r_2,t)$ denotes
the restricted partition sum over all two-line configurations with fixed end
points $(\r_1, \r_2,  t)$, the field
$h(\r_1, \r_2,t) = \beta^{-1} \log Z(\r_1, \r_2,t)$ obeys the equation
\begin{equation}
\partial_t h
= \nu \, \sum_{i=1}^2 \nabla^2_{\r_i} \, h
+ \frac{\lambda}{2} \, \sum_{i=1}^2 ( \nabla_{\r_i} h )^2 \,
+ \sum_{i=1}^2 \eta (\r_i,t)
- g \delta^{d'} (\r_1 - \r_2) \; .
\label{NKPZ}
\end{equation}
Mukherji \cite{Mukherji} uses this
equation to treat the problem of two interacting lines by a standard dynamic
renormalization group approach,  expanding simultaneously in $g$ and
$\lambda$ (see the further remarks below). From (\ref{NKPZ}), one constructs
in a standard way the generating functional of the dynamic correlation
functions \cite{generating.functional},
which are denoted by $\langle \! \langle \dots \rangle \! \rangle$.
This functional integral involves the field $h$ and an auxiliary field
$\widetilde h$. Insertions of the auxiliary field generate response functions,
e.g.
$ \langle \! \langle
  h(\r_1, \r_2, t)
  \prod_{k=1}^{m} \widetilde h (\r_{1k}, \r_{2k}, t_k)
  \rangle \! \rangle =
  \delta^m  \langle \! \langle h(\r_1, \r_2, t) \rangle \! \rangle  /
  \prod_{k=1}^{m} \delta J (\r_{1k}, \r_{2k}, t_k) $,
where \linebreak
$ J (\r_1, \r_2, t) $ is a source term added to the r.h.s. of
(\ref{NKPZ}). By functional differentiation with respect to $J$, one easily
verifies the following relation between connected averages in the polymer
system and dynamic response functions,
\begin{eqnarray}
\beta^{m-1} \;
         \overline{ \langle \delta^{d'} (\r_1 (t_1) - {\bf s}_1)
                            \delta^{d'} (\r_2 (t_1) - {\bf s}_1') \dots
                            \delta^{d'} (\r_1 (t_m) - {\bf s}_m)
                            \delta^{d'} (\r_2 (t_m) - {\bf s}_m')
                    \rangle^c }  & = &  \nonumber
\\
\lim_{T \to \infty}
\langle \! \langle h(\r_1,\r_2,T) \widetilde{h} ({\bf s}_1, {\bf s}_1', t_1)
                     \dots         \widetilde{h} ({\bf s}_m, {\bf s}_m', t_m)
\rangle \! \rangle \quad . \quad\quad\quad\quad  & &
\label{moments}
\end{eqnarray}
The dynamic average
$\langle \! \langle h(\r_1, \r_2 ,t) \rangle \! \rangle $
equals (minus) the disorder-averaged free energy of the polymer system.
For $g = 0$, one has
$\left\langle \! \left\langle h(\r_1, \r_2 ,t)
\right\rangle \! \right\rangle = \sum_{i=1}^2
\left\langle \! \left\langle h(\r_i,t) \right\rangle \! \right\rangle_1 $;
hence $h(\r_1,\r_2,t)$ is a scaling field of dimension $- \omega$ (with time
as the basic scale), as in the usual Kardar-Parisi-Zhang equation. From
(\ref{moments}), we have in particular
$ \overline{ \langle \Psi (t) \rangle } =
  \lim_{T \to \infty}
  \langle \! \langle h(\r_1, \r_2, T) \; \widetilde \psi (t)
  \rangle \! \rangle $
with
$\widetilde \psi (t) \equiv \int {\rm d^{d'}} {\bf r} \;
\widetilde h ({\bf r},{\bf r},t) $.
Comparing dimensions on both sides, we conclude that
$\tilde \psi (t)$ is a dynamic field of dimension $\omega$. Hence its
self-coupling has to be of the form
$ \tilde \psi (t) \tilde \psi (t') \sim  c'
|t - t'|^{- \omega} \tilde \psi (t) + \dots $.
Transforming back to the polymer system then gives (\ref{OPE}).

The renormalization of the perturbation series (\ref{expansion}) to one-loop
order is based on eqns. (\ref{x}) and (\ref{OPE}); it can be borrowed entirely
from ref.~\cite{DEPINNING}. It yields the flow equation
$ L \partial_L u_R = \varepsilon  u_R - c  u_R^2  + O (u_R^3) $
for the renormalized pair coupling $u_R$ (that is proportional to $g$), with
$\varepsilon \equiv 1 - \omega (d')$. To the given order, this equation has
the two fixed points $u_R = 0$ and $u_R^{\ast} = (1 - \omega) / c $. In any
physical dimension $d'$, one has $\varepsilon > 0$. Hence the trivial fixed
point $u_R = 0$ is {\em unstable}, and the interaction is a {\em relevant}
perturbation. This result is at variance \cite{comment.Nattermann} with
ref.~\cite{NattermannAl}.

Any {\em attractive} line-line force ($g < 0$) grows indefinitely in magnitude
under the renormalization. The system develops a {\em bound state} with
transversal  localization length $\xi_\perp$. As $g$ approaches zero from
below, the lines unbind continuously, i.e. the localization length diverges as
\begin{equation}
\xi_\perp \sim \vert g \vert^{- \nu_\perp}
\hspace{1cm}  \mbox{with} \hspace{1cm}
\nu_\perp = \zeta   / \varepsilon =  (1 + \omega)  /  2 (1 - \omega)
\label{xi}
\end{equation}
when the interaction strength  approaches zero.

With a {\em repulsive} interaction ($ g > 0$), however, the large-scale
behavior of the
two-line system is determined by the nontrivial fixed point $u_R^{\ast}$.
At this fixed point, the pair field $\Psi$ acquires the new dimension
$ x^{\ast} = 2 (1 - \omega ) + O( (1 - \omega)^2 )$ that
determines the large-scale asymptotics of the pair distribution function.
For example, the probability of intersection in a system of transversal size
$R$ scales as
\begin{equation}
\overline{ \left\langle \Psi  (t) \right\rangle }
\sim R^{-\theta} \quad \mbox{with} \quad  \theta = x^{\ast} / \zeta \; ,
\label{onepoint.fermi}
\end{equation}
where $\theta \approx 2$ for $d' = 1$ and $\theta \approx 2.4$
for $d' = 2$.
Unlike in the case without pair interactions, the probability of intersection
now approaches zero in the thermodynamic limit $R \to \infty$: the lines avoid
each other completely. Moreover, this implies by standard scaling arguments
that repulsive pair forces have a {\em long-ranged} effect on the pair
distribution function,
\begin{equation}
{\cal P} (\r_1 - \r_2 ) \sim
\vert \r_1 - \r_2 \vert^{ \theta}
\label{pair}
\end{equation}
for $R \gg \vert \r_1 - \r_2 \vert \gg \xi_\perp (g)$.

We may use these results to obtain the finite-size scaling of
the ``overlap''
$q(T, g) \equiv  T^{-1} \int_{0}^{T} {\rm d} t' \,
\overline{ \langle \Psi (t') \rangle }(T, g)$
in a system of {\em longitudinal} length $T$ and with $R \to \infty$,
a quantity that is discussed in the literature \cite{Mezard,Tang,Mukherji}.
We have
$ q(T, g) = q(T, 0) \widetilde Q (T / \xi_{\scriptscriptstyle \|})
  = Q (g \; T^{1 - \omega}) $,
using $ q(T, 0) \sim T^0 $ by (\ref{x})
and the scaling of the logitudinal correlation length
$\xi_{\scriptscriptstyle  \|} \sim \xi_\perp^{1 / \zeta} $ given by (\ref{xi}).
For $d' = 1$, where $1- \omega = 2/3$, this scaling form agrees with
M\'ezard's conjecture \cite{Mezard} (which is based on numerical simulations)
and with Mukherji's renormalization group results \cite{Mukherji}. The latter
approach, however, gives no information on the strong coupling phase in
any dimension $d' > 1$.

In summary, we have shown that directed lines in a highly disordered medium
respond to pair forces in a stronger way than lines with purely thermal
fluctuations. In any dimension $d' < d'_>$, they form a bound state with
attractive forces; with repulsive forces they avoid each other, as described by
the pair distribution function (\ref{pair}). The reason for this strong effect
of short-ranged interactions is that when they are absent, the disorder forces
the lines to cluster in the vicinity of a {\em unique} path of minimal  energy.
Our results hence provide an experimentally and numerically accessible
consequence of the clustering. However, the pair distribution function
(\ref{prob}) shows a {\em singular broadening} for fixed $\tilde
\xi_{\scriptscriptstyle  \|}$ and $R \to \infty$
as $d'$ approaches the upper critical dimension $d'_>$
(i.e. $\omega \searrow 0$): the width $\Delta ( \eta )$ defined by
$\int_{\vert {\bf d} \vert < \Delta} {\rm d^{d'}}{\bf d} \;
{\cal P} (\bf d) = \eta $
diverges as  $\Delta ( \eta ) \sim 1/ \omega$ for any $0 < \eta < 1$, and
accordingly $ \overline{ \langle \Psi \rangle } \sim \omega $.
If $d_>'$ is finite, we hence expect that for $d' > d'_>$, the lines no
longer cluster, but exploit {\em multiple} near-minimal paths even at low
temperature, as in a glassy state.
This leads to a modification of eq.~(\ref{x}),
$ \overline{ \langle \Psi \rangle } \sim R^{-d'} $; weakly attractive pair
forces should then no longer generate a bound state.
This could be useful to  determine $d'_>$ numerically in a way that is less
hampered by finite-size effects than the
existing simulations of the KPZ equation \cite{Ala-NissilaAl}.


\newpage
\begin{thebibliography}{99}
\baselineskip=10pt
\footnotesize{



\bibitem{fluxlines}
M.V. Feigel'man, V.B. Geshkenbein, A.I. Larkin, and V.M. Vinokur,
Phys. Rev. Lett. 63 (1989), 2303;
T. Hwa, Phys. Rev. Lett. 69 (1992), 1552;
D. Nelson, H. S. Seung, Phys. Rev. B 39 (1989), 9153;
D. Nelson, P. Le Doussal, Phys. Rev. B 42 (1990), 10113.


\bibitem{reviews}
G. Forgacs, R. Lipowsky, and T.M. Nieuwenhuizen,  in {\em Phase transitions
and Critical Phenomena}, Vol. 14,  ed. C. Domb and J. Lebowitz
(Academic Press, London, 1991);
M. Kardar, Lectures on Directed Paths in Random Media,
{\em in} Fluctuating Geometries in Statistical Mechanics and Field Theory,
Les Houches proceedings, preprint cond-mat/9411022.


\bibitem{relation.spinglass}
B. Derrida and H. Spohn, J. Stat. Phys. 51 (1988), 817;
M. M\'ezard and G. Parisi, J. Phys. I (France) 1 (1991), 809;
H. Kinzelbach and H. Horner, J. Phys. I (France) 3 (1993), 1329 and 1901.


\bibitem{KPZ}
M. Kardar, G. Parisi, and Y.-C. Zhang, Phys. Rev. Lett.  56 (1986), 889.


\bibitem{KrugSpohn.review}
J. Krug and H. Spohn, in {\em Solids Far From Equilibrium: Growth, morphology
and Defects}, ed. C. Godr\`eche (Cambridge, 1990).


\bibitem{noisy.Burgers}
D.A. Huse, C.L. Henley, and D.S. Fisher, Phys. Rev. Lett. 55 (1985), 2924;
D. Forster, D.R. Nelson, M. Stephen, Phys. Rev. A 16 (1977), 732.


\bibitem{columnar.defect}
M. Kardar, Nucl. Phys. B290 (1987), 582;
L. Balents and M. Kardar, Phys. Rev. B49 (1994) 13030;
L.-H. Tang and I.F. Lyuksyutov, Phys. Rev. Lett. 71 (1993), 2745;
T. Hwa and T. Nattermann, Princeton report, cond-mat 9404031 (1994).


\bibitem{DEPINNING}
H. Kinzelbach, M. L\"assig, Depinning in a Random Medium,
preprint cond-mat/9405088.


\bibitem{Lipowsky.fluctuations}
R. Lipowsky, Europhys. Lett. 15 (1991), 703.


\bibitem{KardarNelson.CIT}
M. Kardar and D.R. Nelson, Phys. Rev. Lett.  55 (1985), 1157.


\bibitem{Mezard}
M. M\'ezard, J. Phys. (France) 51 (1990), 1831.


\bibitem{Tang}
L.-H. Tang, J. Stat. Phys. 77 (1994), 581.


\bibitem{NattermannAl}
T. Nattermann, M. Feigelman, I. Lyuksyutov, Z. Phys. B84 (1991), 353.


\bibitem{Mukherji}
S. Mukherji, Phys. Rev.  E 50 (1994), R2407.


\bibitem{fermions1}
H.B. Thacker, Rev. Mod. Phys. 53 (1981), 253;
M.E. Fisher, J. Stat. Phys. 34 (1984), 667.


\bibitem{fermions2}
S. Mukherji and S.M. Bhattacharjee, J. Phys. A 26 (1993), L1139;
M. L\"assig, Phys. Rev. Lett.  73 (1994), 561.


\bibitem{comment.minpath}
Due to the ``tilt symmetry'', one has
$\overline{ \langle ( \r_i (t) - \r_i (t') )^2  \rangle^c }
= \beta^{-1} \; \vert t - t'  \vert$
exactly.
A degeneracy of ground states in an extensive number of realizations
would lead to temperature-independent corrections to this relation.
See the discussion in
\cite{FisherHuse.paths,HwaFisher.paths} and the results of
\cite{Tang}.


\bibitem{FisherHuse.paths}
D.S. Fisher and D.A. Huse, Phys. Rev. B 43 (1991), 10728.


\bibitem{Parisi}
G. Parisi, J. Phys. (France) 51 (1990), 1595.


\bibitem{HwaFisher.paths}
T. Hwa, D. Fisher, Phys. Rev. B 49 (1994), 3136.


\bibitem{thermalRG}
M. L\"assig and R. Lipowsky, Phys. Rev. Lett. 70 (1993), 1131, discuss a
similar renormalization group for temperature-driven transitions of directed
manifolds.


\bibitem{HuseHenley.roughening}
D.A. Huse and C.L. Henley, Phys. Rev. Lett. 54 (1985), 2708.


\bibitem{MooreAl.modecoupling}
See e.g. M.A. Moore et al., preprint cond-mat/9407076, and references therein.


\bibitem{comment.prob}
$\int {\rm d^{d'}} \! {\bf d} \; {\cal P} ({\bf d}) \; {\bf d}^2  =
\overline{ \langle
\left( \r_1 (t) - \r_2 (t) \right)^2 \; \rangle } =
2 \overline{ \langle
\left( \r(t) \right)^2 \; \rangle^c_1 } =
\lim_{\vert t-t' \vert  \to \infty}  \overline{ \langle
\left( \r(t) - \r(t') \right)^2 \;
\rangle^c_1 }$.
For $\vert t-t' \vert \ll R^{1/\zeta}$, this equals
$ \beta^{-1} \; \vert t - t'  \vert $ due to the tilt
invariance~\cite{FisherHuse.paths,HwaFisher.paths}; in the opposite limit, one
then obtains $\beta^{-1} \;  R^{1/\zeta}$ by finite-size scaling arguments;
see also \cite{HwaFisher.paths}.


\bibitem{generating.functional}
For the general formalism, see e.g.
P.C. Martin, E.D. Siggia, H.A. Rose, Phys. Rev. A8 (1973) 423;
R. Bausch, H.K. Janssen, H. Wagner, Z. Phys. B24 (1976), 113.
Applied to the KPZ-equation, see
T. Sun, M. Plischke, Phys. Rev. E49 (1994), 5046;
E. Frey, U.W. T\"auber, Phys. Rev. E50 (1994), 1024.


\bibitem{comment.Nattermann}
We attribute this discrepancy to the fact that the disorder-induced attraction
between the lines is neglected in the treatment of \cite{NattermannAl}.


\bibitem{Ala-NissilaAl}
T. Ala-Nissila et al., J. Stat. Phys 72 (1993), 207, and references
therein.


}
\end{thebibliography}
\end{document}